\documentclass[onecolumn]{article}

\usepackage{graphicx}
\usepackage{amsmath}
\usepackage{amssymb}
\usepackage{bigstrut}
\usepackage{tabularx,ragged2e,booktabs,caption,authblk}
\newcolumntype{C}[1]{>{\Centering}m{#1}}

\everymath{\displaystyle}
\usepackage{fullpage}
\usepackage{cite}
\usepackage{float}
\usepackage{url}
\usepackage{hyperref}
\usepackage{natbib}
\providecommand{\uppercase}[1]{\MakeUppercase{#1}}

\title{Effective resource allocation to combat invasions of the spotted lanternfly (\textit{Lycorma delicatula}) and similar pests}

\author[]{Daniel Str\"{o}mbom\thanks{\mbox{Corresponding author: stroembp@lafayette.edu}}}
\author[]{Julianna Hoitt}
\author[]{Jinrong Hu}
\author[]{Swati Pandey}
\author[]{Elizabeth Batchelar}
\affil[]{Department of Biology, Lafayette College, Easton, PA 18042, USA}

\date{}

\begin{document}

\maketitle

\begin{abstract}
The spotted lanternfly is rapidly establishing itself as a major insect pest with global implications. Despite substantial management efforts, its spread continues in invaded regions, highlighting the need for refined strategies. A recent model generalized results of empirical control studies by incorporating population dynamics and incomplete delivery, introducing a formula for the minimum proportion of a population that must be treated to induce decline. However, that model cannot address the more practical question of how to allocate control efforts. Here, we extend the model to identify effective deployment strategies. When control effects scale linearly with effort, we show that sequential application of stage‑specific controls, ordered by efficacy, is optimal. When effects exhibit diminishing returns, we derive a switching criterion between controls that accommodates variable or uncertain resources. For fixed resources, we employ global optimization to obtain deployment strategies. Both approaches consistently outperform random deployment, which we show can lead to catastrophic outcomes. Our findings demonstrate the importance of adopting effective control strategies for the lanternfly. Despite this need, we found no prior studies addressing deployment strategies in the lanternfly literature. The methods developed here provide a foundation for ensuring that limited management resources are used effectively.
\end{abstract}

\section*{Introduction}

The spotted lanternfly (\textit{Lycorma delicatula}) is an invasive insect pest that has spread beyond its native range in recent years, establishing a presence in South Korea, Japan, and the United States \citep{Dara2015,Barringer2015}. Several modeling studies based on climate and other factors also suggest that the spotted lanternfly may emerge as a global invasive pest, with predicted regions of infestation on all continents except Antarctica \citep{Jung2017,Wakie2020}. In infested regions, it is causing damage to both economically important farmed plants and naturally occurring vegetation, making it a management priority both where it is established \citep{Urban2020} and in preventing its establishment in new regions \citep{EPPO}. Unfortunately, in many of the recently invaded areas, it lacks effective natural enemies to help control the population \citep{Johnson2023,Strombom2024a}, making human intervention necessary \citep{Aphis2018}. In the U.S., quarantine zones have been established, and a variety of control measures have been implemented, including insecticides \citep{Leach2019,Shin2010}, entomopathogens \citep{Clifton2020}, egg parasites \citep{Liu2019a,Wu2023}, traps \citep{Francese2020}, egg scraping \citep{Cooperband2018}, removal of its preferred host plant \citep{Urban2020}, and even instructing the public to kill lanternflies on sight \citep{Urban2023}. Despite these efforts, the lanternfly has continued to spread from Berks County, Pennsylvania, where it was first observed in 2014, to numerous counties across 18 states, and the spread shows no signs of slowing down \citep{NYSIPM2024,Cook2021,Jones2022,Strombom2024b}, suggesting that current management efforts are insufficient.

Recently, a comparative modeling study of the efficacy of several proposed, and used, control measures when population dynamics and incomplete delivery are taken into account provided a plausible partial explanation for the observed lack of effect on the growth and spread of the US lanternfly population \citep{Strombom2021}. Previous empirical studies of control measure effectiveness suggested that several of the control measures may be effective in combating the lanternfly due to the significant mortality rates observed in the studies \citep{Clifton2020,Cooperband2018,Cooperband2019}. However, when population dynamics and the possibility of incomplete delivery of the control are introduced, the expected effectiveness of all controls was reduced. More specifically, they introduce the generalized population growth formula (Equation 4 in \citep{Strombom2021})
\begin{equation}\label{eq:1} \lambda^x(p)=[ps_{PO}^x+(1-p)s_{PO}][ps_{NP}^x+(1-p)s_{NP}][ps_{AN}^x+(1-p)s_{AN}][pF^x+(1-p)F], \end{equation}
where $s_{PO}$, $s_{NP}$, $s_{AN}$, and $F$ are the natural survival and reproduction parameters and $s^x_{PO}$, $s^x_{NP}$, $s^x_{AN}$, and $F^x$ the corresponding survival and reproduction parameters when a control $x$ is applied, and $p$ is the proportion of lanternfly treated. By parameterizing this formula using data from the literature (see Table 1 in \cite{Strombom2021} for stage-specific survival rates under different control measures), they found that when population dynamics and incomplete delivery are accounted for, even a perfect control that kills all treated SLF (e.g. some insecticides) requires at least 35\% of all lanternfly in each stage to be treated to turn population growth into decline. For less-than-perfect controls (e.g. entomopathogens \citep{Clifton2020} and egg parasites \citep{Liu2019a}), an even higher treated proportion is required for population decline, if possible at all (see Figure 2 in \citep{Strombom2021}).

However, while \citep{Strombom2021} established an estimate of the annual population growth rate, the expected effectiveness of a number of control measures and combinations of them, and an absolute lower bound on the proportion that needs to be treated in each stage (0.35) numerically, their work provides no information about how to deploy a set of controls effectively. For example, if a number of control measures with specified efficacies are available, how should they be deployed for maximal effect? Should the available resources be allocated equally across all life stages, should we concentrate the efforts on one life stage, or use some other resource allocation scheme? To address questions of this type, a more sophisticated approach is required. In particular, one that distinguishes the proportion treated in the different life-stages, rather than having a common proportion treated for all, given that the available controls have different stage-specific efficacies. For example, some controls only affect one stage, e.g. egg parasites \citep{Liu2019a}, mid-winter chipping \citep{Cooperband2018}, insecticide Chlorpyrifos \citep{Shin2010,Leach2019}, whereas others affect multiple stages but may have different efficacy depending on the exact stage, e.g. entomopathogen \textit{B. bassiana} \citep{Clifton2020} and insecticide Bifenthrin \citep{Leach2019}.

Furthermore, in addition to differences in efficacy of a control once delivered to the insect, the process of delivering the control to the population may exhibit diminishing returns with effort \citep{Oerke2006,Mitchell2014,Brown2017,Edholm2018} and possibly other non-linear relationships between stage-specific proportion treated/effort expended and reduction in population growth. We note that the framework introduced in \citep{Strombom2021} is based on the assumption that the effect of applying a control grows linearly with proportion treated/effort expended, i.e. effect is proportional to effort applied, and it will therefore have to be modified to accommodate any non-linear response such as diminishing returns.

The development and use of optimization and effectivization tools to address questions of how to deploy or allocate resources are ubiquitous across science and technology \citep{Diwekar2020}, and are also common in research on invasive insect pest control, e.g. \citep{Yemshanov2017,Edholm2018,Buyuktahtakin2018,Thompson2021}. These sources illustrate the utility of general optimization frameworks for pest management, however, our approach differs in that it builds directly on a species-specific life-cycle model tailored to the lanternfly \citep{Strombom2021}, rather than adapting a general optimization model to this system. To our knowledge, no work in this direction exists for the spotted lanternfly. The main contribution of this paper is twofold: (1) we generalize the incomplete delivery model in \citep{Strombom2021} to allow stage-specific allocation under diminishing returns, and (2) we use this framework to formulate and analyze optimization problems that provide guidance for effective resource allocation. Work of this type may be critical for informing management efforts, both in areas where the lanternfly is currently present and in regions globally where it may appear in the future. In already infested regions, such models could serve as tools for the effective allocation of resources to manage populations. For future invasions, if detected early enough, this information could potentially contribute to eradication. 

\section*{Results}
In this section we present the main results only and detailed derivations are given in Methods and Calculations (Appendix). The results are organized as a sequence of increasingly general problem formulations, beginning with the population growth model introduced in \citep{Strombom2021}. First, we re-express this model in a more compact form and then extend it to allow stage-specific proportions treated, which makes it possible to compare the relative effectiveness of different stage-specific controls. Next, we consider the case where stage-specific effects exhibit diminishing returns, reformulating the problem in terms of effort variables $\epsilon_i$ and deriving conditions for switching between controls. We then apply numerical global optimization and characterize random deployment outcomes. In Methods and Calculations (Appendix) we also provide a numerical optimization protocol to identify effective allocation strategies for the fully general situation where the functional relationships $p_i(\epsilon_i)$ are unknown. This progression illustrates how successive generalizations of the model allow increasingly realistic and practically useful insights into resource allocation.

\subsubsection*{Result 1}
The original population growth function (\ref{eq:1}) in terms of the proportion treated, $p$, can be expressed as
\begin{equation}\label{eq:2} \lambda^x(p) = \lambda_0 \prod_{i=1}^4 \left(1 - p(1 - k_i)\right), \end{equation}
if we define $k_1 = s^x_{PO}/s_{PO}$, $k_2 = s^x_{NP}/s_{NP}$, $k_3 = s^x_{AN}/s_{AN}$, $k_4 = F^x/F$, and $\lambda_0 = s_{PO} s_{NP} s_{AN} F$.
Furthermore, replacing the common proportion $p$ used for all stages in (\ref{eq:2}) with stage-specific proportions treated, $\bar{p} = [p_1, p_2, p_3, p_4]$, yields
\begin{equation}\label{eq:3} \lambda^x(\bar{p}) = \lambda_0 \prod_{i=1}^4 \left(1 - p_i(1 - k_i)\right), \end{equation}
with $p_i \in [0,1]$ and $k_i \in [0,1)$, where the non-inclusive 1 endpoint comes from the assumption that all controls have some effect. A control with no effect corresponds to $s_i^x = s_i$ so that $k_i = 1$, in which case that stage simply grows at its natural rate and allocation to it has no influence on the results.

We note that empirical values of the stage-specific parameters $s_i$ and $s_i^x$ are reported in (\citep{Strombom2021}, Table 1), from which $k_i$ can be calculated for several empirically studied controls.

See Methods and Calculations (Appendix), Result 1, for more details.

\subsubsection*{Result 2}
To minimize the growth rate in situations where the effect of an applied stage-specific control is proportional to the effort expended, as in (\ref{eq:3}), the available resources should be allocated to sequentially maximize the stage-specific proportions treated in the order specified by the effectiveness of the control on the stages. That is, maximize resource allocation to the $p_i$ corresponding to the smallest $k_i$ first. If there are resources left, proceed to maximize the $p_j$ corresponding to the second smallest $k_j$, and so on. This follows from the fact that if $k_i < k_j$ (and $p_i > p_j\frac{1 - k_j}{1 - k_i}$), then
\begin{equation}\label{eq:4} \frac{\partial\lambda^x}{\partial p_i} < \frac{\partial\lambda^x}{\partial p_j}. \end{equation}
Intuitively, this means that a stage with a smaller $k_i$ (more effective control) should always be prioritized, since allocating effort there yields a larger marginal reduction in growth than allocating the same effort to a less effective stage. See Methods and Calculations (Appendix), Result 2, for a numerical example. Note that the condition $p_i > p_j\frac{1 - k_j}{1 - k_i}$ is not restrictive, because $\frac{1 - k_j}{1 - k_i} < 1$, and it is trivially satisfied if we begin by increasing the proportion treated corresponding to the most effective stage-specific control $i$ (so $p_i > 0$ and $p_j = 0$).

See Methods and Calculations (Appendix), Result 2, for more details.

\subsubsection*{Result 3}
\begin{figure}[h!] \includegraphics[width=\columnwidth]{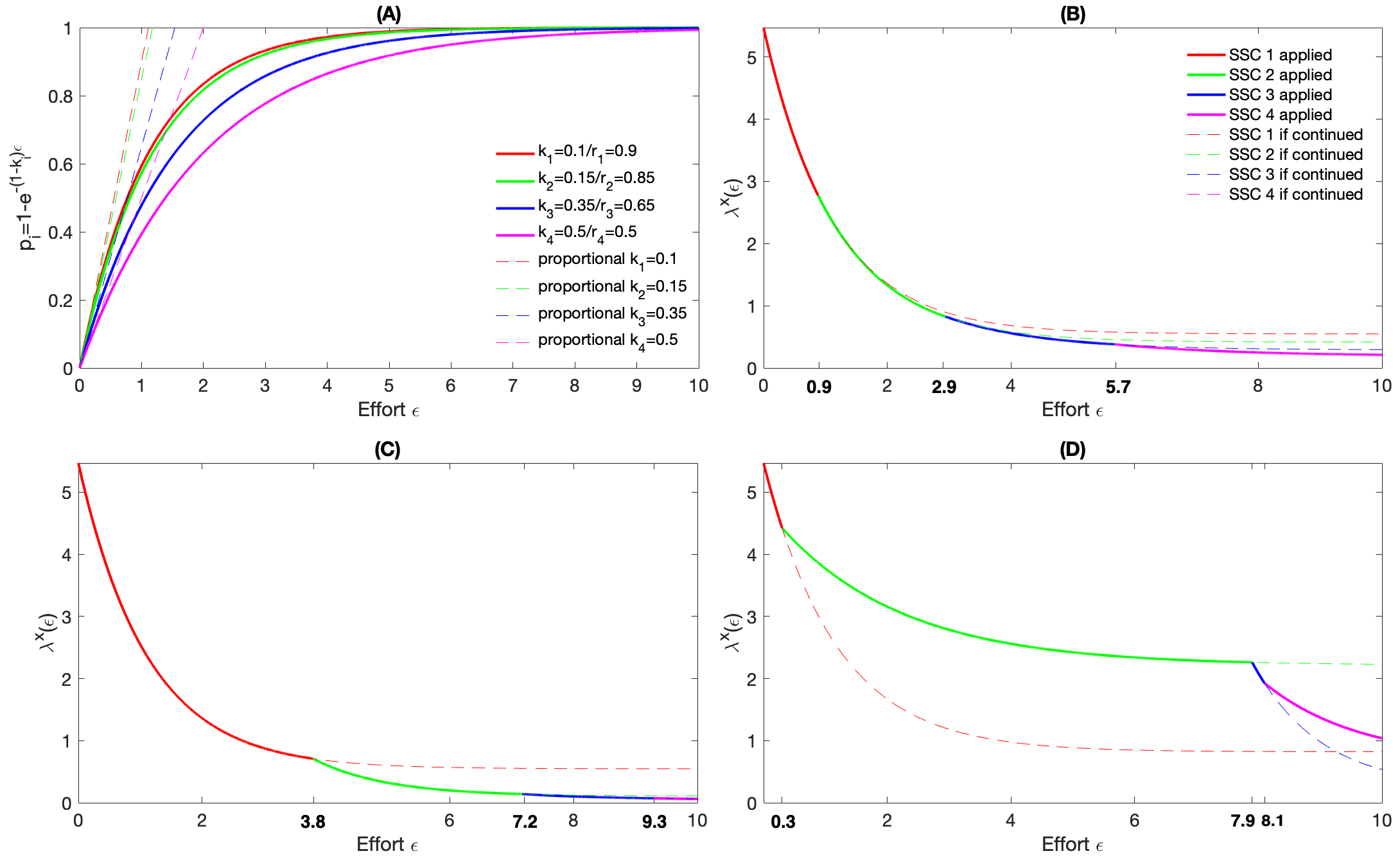} \caption{{\bf} Illustration of proportion treated curves and outcomes under different stage-specific control (SSC) strategies. 
(A) Example proportion treated curves in the diminishing returns case, with solid curves showing asymptotic behavior and dashed lines representing linear approximations. 
(B) Population growth rate when SSC efforts are allocated according to formula (\ref{eq:9}); switching between SSCs occurs at efforts $\epsilon=0.9$, $2.9$, and $5.7$ (solid vs. dashed curves indicate improved outcomes when switching). 
(C) Population growth rate when SSC efforts are allocated by numerical optimization algorithms, with switching at $\epsilon=3.8$, $7.2$, and $9.3$ (again, solid vs. dashed curves indicate improved outcomes when switching). 
Comparing (B) and (C), the formula strategy yields lower growth rates for $\epsilon$ between 0.9 and 3.8, while optimization yields the best final result at $\epsilon=10$. 
(D) Example outcome when SSC order and effort allocation are random, leading to poorer performance (final growth rate $>1$). For illustration we use the empirically estimated annual growth rate $\lambda_0=5.47$ (from \citep{Strombom2021}) together with a representative set of realistic but hypothetical $k_i$ values chosen to best demonstrate the method, while noting that the framework itself is compatible with any empirically derived controls. All curves are obtained by plotting the corresponding functions or by simulations/optimizations as described in Methods and Calculations (Appendix) (with full MATLAB code available, see Code Availability Statement).} \label{fig:1} \end{figure}

In situations where the effect of applying effort into a stage-specific control exhibits diminishing returns, the relative efficacies of the stage-specific controls are not absolute and will depend on the effort already expended. Let $\epsilon_i \in [0, \infty)$ be the effort expended on deploying control in stage $i$, and define the proportion treated $p_i$ to be the standard diminishing returns function $p_i=1-e^{-r_i\epsilon_i}$, where $r_i > 0$ is a constant associated with the effectiveness of the control when applied to stage $i$.  
In this case, the population growth formula (\ref{eq:3}) becomes
\begin{equation}\label{eq:5} \lambda^x(\bar{\epsilon})=\lambda_0 \prod_{i=1}^4(1-(1-e^{-r_i\epsilon_i})(1-k_i)). \end{equation}
In this formulation, $r_i$ is the initial marginal gain in coverage per unit effort ($dp_i/d\epsilon_i$ at $\epsilon_i=0$). To make the diminishing-returns case match the proportional case for $\epsilon_i=0$, we set $r_i=1-k_i$, which allows direct comparability of the analytic results, and practical estimation of $r_i$ from data is described in Methods and Calculations (Appendix), Result 3. See Figure \ref{fig:1}A for example stage-specific proportion treated curves in the diminishing returns case, where solid curves show the diminishing-returns curves exhibiting asymptotic behavior and the corresponding dashed curves show the proportional case as linear approximations at $\epsilon=0$. With this function, the corresponding population growth rate formula for the diminishing returns case is
\begin{equation}\label{eq:7} \lambda^x(\bar{\epsilon})=\lambda_0 \prod_{i=1}^4(1-(1-e^{(k_i-1)\epsilon_i})(1-k_i)) \end{equation}
with $\epsilon_i\in[0,\infty)$ and $k_i\in[0,1)$, or in terms of $r_i$,
\begin{equation}\label{eq:8} \lambda^x(\bar{\epsilon})=\lambda_0 \prod_{i=1}^4(1-r_i(1-e^{-r_i\epsilon_i})). \end{equation}

See Methods and Calculations (Appendix), Result 3, for more details.

\subsubsection*{Result 4}
If we order the stage-specific controls by initial efficacy (at $\epsilon_i=0$), i.e. $r_i > r_{i+1}$, and expend effort $\epsilon_i$ into the initially most effective $i$:th control, then this $i$:th control will become less effective in reducing growth than the initially second most effective $i+1$:th control at $\epsilon_{i+1}=0$ when $\epsilon_i$ reaches the value given by
\begin{equation}\label{eq:9} \epsilon_{i}=\frac{1}{r_i}\ln\frac{r_i(r_i-r_{i+1}^2)}{r_{i+1}^2(1-r_i)}. \end{equation}
Therefore, switching to apply effort into the $i+1$:th control at this effort value will lead to an improvement compared to continuing to expend effort on control $i$. See Figure \ref{fig:1}B, which illustrates the population growth rate when resources are allocated according to the switching-effort formula (\ref{eq:9}), with solid vs. dashed curves showing the improved outcome achieved by switching at the calculated effort thresholds.. This formula is valid for all $r_i, r_{i+1} \in (0,1)$, which is the relevant range, as long as $r_i > r_{i+1}$.

See Methods and Calculations (Appendix), Result 4, for more details.

\subsubsection*{Result 5}
While using the switching time formula will lead to a larger reduction in growth rate than continuing to deploy previously more effective controls or randomly deploying controls (Figure \ref{fig:1}D), it does not necessarily yield the optimal result in situations with a fixed budget constraint. In such cases, numerical global optimization methods may produce solutions that achieve a greater reduction in growth rate for the same amount of effort. See Figure \ref{fig:1}C , which shows the population growth rate when resources are allocated using numerical global optimization, with switches occurring at efforts $\epsilon=3.8, 7.2,$ and $9.3$, while Figure \ref{fig:1}D shows an example of random deployment of controls that switches at $\epsilon=0.3, 7.9,$ and $8.1$ and yields a final growth rate larger than 1.
For the parameters corresponding to the example in Figure \ref{fig:1}, random deployment of controls yields an average final growth rate of 0.4, compared to 0.2 when formula (\ref{eq:9}) is used, and 0.06 when global optimization is used. See Methods and Calculations (Appendix), Result 5, for details about how these values were obtained. Thus, formula (\ref{eq:9}) is about twice as effective at reducing the growth rate, and optimization about seven times as effective, compared to the average random deployment of controls. We also note that the maximum final growth rate for random deployment of controls is 2.75, which is about 14 times less effective than using formula (\ref{eq:9}) and about 46 times less effective than using optimization. However, it is important to note that if effort is cut prematurely, for example at $\epsilon=3$ in Figure \ref{fig:1}, then formula (\ref{eq:9}) would yield a better final result than the global optimization. Thus, the global optimization strategy consistently yields the best final outcomes when total effort is high, while the switching-time formula provides better results at intermediate effort levels, and random deployment performs worst overall.

See Methods and Calculations (Appendix), Result 5, for more details.

\section*{Discussion}
The overall conclusion of this work is that adopting a suitable strategy for the deployment of stage-specific controls is essential for effective management. In particular, random deployment of controls may be inefficient and can lead to disastrous outcomes. Care should therefore be taken when deciding whether to opt for a "greedy" strategy, such as formula (\ref{eq:9}), or a strategy based on numerical global optimization results. For a fixed guaranteed amount of resources, the numerical optimization strategy is always preferable. However, if resources are not fixed and may be subject to change, a strategy based on formula (\ref{eq:9}) may be preferable. We find that considerations of this type are largely absent from the quantitative pest control literature that focuses on using optimization approaches (e.g. those in \citep{Buyuktahtakin2018,Thompson2021}), despite the fact that resources often are subject to change \citep{Ahmed2022}. This suggests that many methods that are marketed as optimal may produce suboptimal results in practice, unless the problem has an optimal substructure. This distinction is also useful to communicate to funders of, and agencies responsible for executing, management efforts, namely, that the same amount of resources will yield better results if provided through a stable, long-term program than if they are provided on a more unreliable, short-term basis and subject to potential change. Delays and lags in management efforts resulting from budgetary or urgency-of-action considerations are known to adversely impact the prevention and eradication of invasive species \citep{Ahmed2022}. For example, the management strategy of Berks County, PA, was set back by a loss of state funding in 2021, despite a continued threat from the spotted lanternfly to the agricultural industry and businesses in the county \citep{Scheid2021}. In 2017, in Pennsylvania alone, the spotted lanternfly was projected to cause $42.6$ million in annual damages. Maintaining management practices to slow the spread could help minimize the risk of economic damage to both local and federal governments \citep{Harper2019}.

The re-written version (\ref{eq:2}) of the original generalized growth formula (\ref{eq:1}) from \citep{Strombom2021} makes it easier to recognize that an underlying assumption of the original model is that the effect of an applied control is proportional to the effort expended. While this might be a reasonable assumption in some situations, such as when managing lanternfly populations in a small, limited geographical area, it is likely that diminishing returns will be substantial in many real-world scenarios \citep{Oerke2006,Mitchell2014,Brown2017,Edholm2018}. Identifying this assumption and developing approaches that incorporate diminishing returns is therefore critical, not only for the application of the methods introduced here, but also for extending the work in \citep{Strombom2021} and other influential models of invasive species control across taxa, e.g. \citep{Hastings2006,Blackwood2010,Moody1988,Buyuktahtakin2018,Thompson2021}. Here, we use the exponential form $p_i = 1 - e^{-r_i \epsilon_i}$ because negative exponential "kill functions" of the form $e^{-r_i \epsilon_i}$ are widely used in pest management models to represent pesticide efficacy and residual decay \citep{Bor1995,Liang2012, Liu2021}. Our $p_i$-function simply reformulates this in terms of the proportion treated and yields the same saturating effort–response behavior. Other functional forms, such as logistic or power-law, may also be useful to investigate for specific applications.

Here, we have only considered the switching time formula (\ref{eq:9}) in the range that is appropriate for our specific application, i.e. with $k_i \in (0,1)$ and $r_i = 1 - k_i \in (0,1)$. However, it is worth noting that formula (\ref{eq:9}) has a region of validity for $r_i > 1$ as well, which may be useful for other applications involving diminishing returns, for example, if the $r_i$ represent rates that can exceed 1, which is common. In such cases, formula (\ref{eq:9}) can be used to determine switching times in the same way we have used it here, as long as $r_i > r_{i+1} > \sqrt{r_i}$ for all $i$.

For practical purposes, the function $p_i = p_i(\epsilon_i)$ will, in general, be unknown, and biological data and arguments should be used to estimate it based on the specific biological situation. It may also be the case that, in practice, each stage has a different such function. To handle situations like these, numerical methods will often be required. We used the Mathematica function {\tt Minimize} \citep{Mathematica} to obtain the global optimization results here and in Methods and Calculations (Appendix) we explain how the command we used can be modified to perform the corresponding optimization in a number of different situations, including different stage-specific $p_i = p_i(\epsilon_i)$ functions. For example, this includes logistic or power-law functions such as those mentioned above.

We also note that for some applications or questions a particular stage might not be affected by a control. That would amount to $s^x_i=s_i$ and thus $k_i=1$. By definition, a genuine control should reduce survival or reproduction at least to some extent. If $k_i=1$ for a given stage, then that stage contributes only its natural growth rate and cannot be affected by treatment. In such cases, the corresponding terms drop out of the optimization problems, and no switching efforts or allocation rules are associated with that stage.

Given that the resources available to combat the lanternfly are limited, we should strive to ensure that they are used effectively. The work introduced here, as well as other optimization approaches that are used regularly in other areas \citep{Diwekar2020,Yemshanov2017,Edholm2018}, can contribute to this goal. The fact that we were unable to find any information on the allocation strategies used for lanternfly management, nor any studies focusing on evaluating such strategies, is concerning. In part, this raises the possibility that some kind of heuristic such as "we use what we have for whatever action seems most urgent or convenient at the time" may, in fact, be used in practice. This would amount to an instance of random deployment, as we have defined it here, and as we have shown, is likely to be suboptimal with respect to the effective use of resources and may lead to disastrous management outcomes.

\section*{Appendix. Methods and Calculations}

Here we present the main calculations, derivations, and methods used to obtain the results.

\subsection*{Result 1}
The generalized population growth formula presented as Equation 4 in \citep{Strombom2021} is
\begin{equation}\label{eq:10} \lambda^x(p) = [p s_{PO}^x + (1 - p) s_{PO}][p s_{NP}^x + (1 - p) s_{NP}][p s_{AN}^x + (1 - p) s_{AN}][p F^x + (1 - p) F]. \end{equation}
First, we replace the common $p$ in this equation with individual proportions for each stage, $p_{PO}$, $p_{NP}$, $p_{AN}$, and $p_{F}$, which yields

\begin{equation}\label{eq:11}
\begin{aligned}
\lambda^x(\bar{p}) &= [p_{PO} s_{PO}^x + (1 - p_{PO}) s_{PO}] \\
&\quad \times [p_{NP} s_{NP}^x + (1 - p_{NP}) s_{NP}] \\
&\quad \times [p_{AN} s_{AN}^x + (1 - p_{AN}) s_{AN}] \\
&\quad \times [p_{F} F^x + (1 - p_{F}) F].
\end{aligned}
\end{equation}
Next, we express the parameter values under control as proportions of the natural parameter values, i.e. $s_{PO}^x = k_{PO} s_{PO}$, $s_{NP}^x = k_{NP} s_{NP}$, $s_{AN}^x = k_{AN} s_{AN}$, and $F^x = k_{F} F$, which gives
\begin{equation}\label{eq:12} \lambda^x(\bar{p}) = \lambda_0 (1 - p_{PO}(1 - k_{PO}))(1 - p_{NP}(1 - k_{NP}))(1 - p_{AN}(1 - k_{AN}))(1 - p_{F}(1 - k_{F})), \end{equation}
where $\lambda_0 = s_{PO} s_{NP} s_{AN} F$ is the natural annual growth rate (Equation 2 in \citep{Strombom2021}).
Finally, if we replace the subscript notation with indices $i=1,2,3,4$, the expression becomes
\begin{equation}\label{eq:13} \lambda^x(\bar{p}) = \lambda_0 \prod_{i=1}^4 (1 - p_i(1 - k_i)), \end{equation}
where $k_i = s^x_i / s_i$. We note that the variables satisfy $p_i \in [0,1]$, since the $p_i$ represent proportions treated, and $k_i \in [0,1)$, where the non-inclusive upper bound follows from the assumption that all controls have some effect.

\subsection*{Result 2}
Result 2 states that if $k_i < k_j$ and $p_i > p_j \frac{1 - k_j}{1 - k_i}$, then
\begin{equation} \frac{\partial \lambda^x}{\partial p_i} < \frac{\partial \lambda^x}{\partial p_j}. \end{equation}
To show this, define $\alpha_i = 1 - p_i(1 - k_i)$ so that (\ref{eq:3}) can be written as
\begin{equation} \lambda^x(\bar{p}) = \lambda_0 \prod_{i=1}^4 \alpha_i, \end{equation}
and note that the derivative of $\lambda^x(\bar{p})$ with respect to $p_i$ is given by
\begin{equation} \frac{\partial \lambda^x}{\partial p_i} = -\lambda_0 (1 - k_i) \prod_{j \neq i} \alpha_j. \end{equation}   
Then we have:
\begin{align} \frac{\frac{\partial \lambda^x}{\partial p_i}}{\frac{\partial \lambda^x}{\partial p_j}}&= \frac{\lambda_0 (1 - k_i) \prod_{k \neq i} \alpha_k}{\lambda_0 (1 - k_j) \prod_{k \neq j} \alpha_k}
= \frac{(1 - k_i) \alpha_j}{(1 - k_j) \alpha_i} \ = \left( \frac{1 - k_i}{1 - k_j} \right) \left( \frac{1 - p_j(1 - k_j)}{1 - p_i(1 - k_i)} \right)\\
&> 1 \cdot \left( \frac{1 - p_j(1 - k_j)}{1 - p_i(1 - k_i)} \right) > 1, \end{align}
where the first inequality holds because $k_i < k_j$ and the second holds when $p_i > p_j \frac{1 - k_j}{1 - k_i}$. Therefore,
\begin{equation} \frac{\frac{\partial \lambda^x}{\partial p_i}}{\frac{\partial \lambda^x}{\partial p_j}} > 1, \end{equation}
which implies that
\begin{equation} \frac{\partial \lambda^x}{\partial p_i} < \frac{\partial \lambda^x}{\partial p_j}, \end{equation}
where the switch in inequality direction results from the fact that the partial derivatives are strictly negative.
We also note that, when formulated in more general terms, this is a well-known result typically established via linear programming \citep{Dantzig1963}.

\subsubsection*{Example numerical illustration of Result 2}
With $\lambda^x(\bar{p}) = \lambda_0 \prod_{i=1}^4 (1 - p_i(1 - k_i))$, the partial derivatives are
$\frac{\partial \lambda^x}{\partial p_i} = -\lambda_0\\ (1 - k_i) \prod_{j \neq i} (1 - p_j(1 - k_j)).$ At $\bar{p}=\bar{0}$ this reduces to $\frac{\partial\lambda^x}{\partial p_i}\big|_{\bar p=\mathbf{0}}=-\lambda_0(1-k_i)$, so the stage with the smallest $k_i$ yields the largest marginal reduction in $\lambda^x$. Using the values corresponding to the example in Figure \ref{fig:1}, i.e. $k_1=0.10$, $k_2=0.15$, $k_3=0.35$, $k_4=0.50$ gives\\

$\frac{\partial\lambda^x}{\partial p_1}=-0.90\lambda_0,\quad
\frac{\partial\lambda^x}{\partial p_2}=-0.85\lambda_0,\quad
\frac{\partial\lambda^x}{\partial p_3}=-0.65\lambda_0,\quad
\frac{\partial\lambda^x}{\partial p_4}=-0.50\lambda_0,$\\

\noindent hence $|\partial\lambda^x/\partial p_1|>|\partial\lambda^x/\partial p_2|>|\partial\lambda^x/\partial p_3|>|\partial\lambda^x/\partial p_4|$, which matches Result 2: prioritize the smaller $k_i$ first. The same ordering holds for nonzero $\bar p$ whenever $p_i>p_j\frac{1-k_j}{1-k_i}$, as stated in the result.

\subsection*{Result 3}
To see that for the diminishing returns case, $p^d(\epsilon) = 1 - e^{-r_i \epsilon}$, to match the proportional case, $p^p(\epsilon) = \epsilon(1 - k_i)$, at $\epsilon=0$, we must have $r_i = 1 - k_i$, note that
\begin{equation} \frac{dp^d}{d\epsilon}\bigg|_{\epsilon=0} = r_i e^{-r_i \cdot 0} = r_i \end{equation}
and
\begin{equation} \frac{dp^p}{d\epsilon}\bigg|_{\epsilon=0} = 1 - k_i, \end{equation}
so for them to match, we require $r_i = 1 - k_i$.\\

In empirical applications, however, $r_i$ need not be set equal to $1-k_i$. Instead, $r_i$ can be estimated directly from effort–coverage data by fitting the function $p_i(\epsilon_i) = 1 - e^{-r_i \epsilon_i}$. For a single observation $(\epsilon_i, p_i)$ we can find $r_i$ through $r_i = -\ln(1-p_i)/\epsilon_i$, and for multiple observations $r_i$ can be obtained by least-squares regression. This provides a direct link between field data (e.g. trap density, spray volume, or labor effort ($\epsilon_i$) and proportion affected ($p_i$)) and the parameter $r_i$.

\subsection*{Result 4}
To derive formula (\ref{eq:9}), define $\sigma_i(\epsilon_i) = 1 - r_i(1 - e^{-r_i\epsilon_i})$ so that (\ref{eq:8}) can be written as
\begin{equation} \lambda^x(\hat{\epsilon}) = \lambda_0 \prod_{i=1}^4 \sigma_i(\epsilon_i) \end{equation}
and note that the derivative of $\lambda^x(\hat{\epsilon})$ with respect to $\epsilon_i$ is given by
\begin{equation} \frac{\partial \lambda^x}{\partial \epsilon_i} = \lambda_0 r_i^2 e^{-r_i\epsilon_i} \prod_{j \neq i} \sigma_j(\epsilon_j). \end{equation}
Assume that the controls are ordered sequentially by effectiveness, i.e. $k_i < k_{i+1}$, or equivalently $r_i > r_{i+1}$, or equivalently $\sigma_i(0) > \sigma_{i+1}(0)$. To find the $\epsilon_i$ at which further increasing $\epsilon_i$ will yield a lower reduction in population growth rate than starting to put effort $\epsilon_{i+1}$ into the initially less effective control, we solve 
\begin{equation} \frac{\partial \lambda^x}{\partial \epsilon_i} = \frac{\partial \lambda^x}{\partial \epsilon_{i+1}}\bigg|_{\epsilon_{i+1}=0}, \end{equation}
\begin{equation} r_i^2 e^{-r_i\epsilon_i} \prod_{j \neq i} \sigma_j(\epsilon_j) = r_{i+1}^2 \prod_{j \neq i+1} \sigma_j(\epsilon_j), \end{equation}
\begin{equation} r_i^2 e^{-r_i\epsilon_i} \sigma_{i+1}(0) = r_{i+1}^2 \sigma_i, \end{equation}
\begin{equation} r_i^2 e^{-r_i\epsilon_i} \left(1 - r_{i+1}(1 - e^{-r_{i+1}0})\right) = r_{i+1}^2 \left(1 - r_i(1 - e^{-r_i\epsilon_i})\right), \end{equation}
\begin{equation} r_i^2 e^{-r_i\epsilon_i} = r_{i+1}^2 \left(1 - r_i(1 - e^{-r_i\epsilon_i})\right), \end{equation}
\begin{equation} \frac{r_i^2}{r_{i+1}^2} = e^{r_i\epsilon_i} \left(1 - r_i\right) + r_i, \end{equation}
\begin{equation} e^{r_i\epsilon_i} \left(1 - r_i\right) = \frac{r_i^2}{r_{i+1}^2} - r_i, \end{equation}
\begin{equation} e^{r_i\epsilon_i} = \frac{r_i^2 - r_i r_{i+1}^2}{r_{i+1}^2(1 - r_i)} = \frac{r_i(r_i - r_{i+1}^2)}{r_{i+1}^2(1 - r_i)}, \end{equation}
\begin{equation} \epsilon_i = \frac{1}{r_i} \ln \frac{r_i(r_i - r_{i+1}^2)}{r_{i+1}^2(1 - r_i)}. \end{equation}
This formula is valid for all $r_i, r_{i+1} \in (0,1)$, which is the relevant range for our application, as long as the controls are ordered by initial efficacy so that $r_i > r_{i+1}$. This follows from the fact that the logarithm part of the formula is well-defined for all $r_i, r_{i+1} \in (0,1)$ such that $r_i - r_{i+1}^2 > 0$, which is equivalent to $r_{i+1} < \sqrt{r_i}$. For $r_i \in (0,1)$, it holds that $r_i < \sqrt{r_i}$, i.e. the square root of $r_i$ is larger than $r_i$, and since we assume that $r_i > r_{i+1}$, it automatically holds that $r_{i+1} < \sqrt{r_i}$ for all $r_i, r_{i+1} \in (0,1)$.
Furthermore, although it is beyond the scope of this paper, we note that the formula has a region of validity for $r_i > 1$ as well. For the formula to be well-defined in this case, $r_i - r_{i+1}^2 < 0$ must hold, which implies $r_{i+1} > \sqrt{r_i}$. For $r_i > 1$, it holds that $r_i > \sqrt{r_i}$, i.e. the square root of $r_i$ is smaller than $r_i$, and since we assume $r_i > r_{i+1}$, the formula is also valid for all $r_i > r_{i+1} > \sqrt{r_i}$.

\subsection*{Result 5}
\subsubsection*{Random application of controls}
To characterize random deployment of controls for efficiency comparisons with the switch time formula and numerical optimization, we implemented code in MATLAB that randomizes both the order of the stage-specific control efficacies ($k_i$) and the effort deployed into each ($\epsilon_i$) in (\ref{eq:7}) over $10^6$ runs, collecting the final growth rate in each. We then calculated the mean (0.40), standard deviation (0.42), minimum (0.06), and maximum (2.75) of this set. See the Code Availability Statement for information on how to access the complete code.

\subsubsection*{Global optimization}
To perform the constrained global numerical optimization we used the Mathematica function {\tt Minimize} \citep{Mathematica}. For the example in Figure 1 the following command was used {\tt Minimize[\{5.47*(1 - 0.9*(1 - Exp[-0.9 x]))*(1 - 0.85 * (1 - Exp[-0.85 y]))*(1 - 0.65*(1 - Exp[-0.65 z]))*(1 - 0.5*(1 - Exp[-0.5 w])), x + y + z + w == 10 \&\& x >= 0 \&\& y >= 0 \&\& z >= 0 \&\& w >= 0\},  \{x, y, z, w\}]} where $x$, $y$, $z$ and $w$ are placeholders for $\epsilon_1$, $\epsilon_2$, $\epsilon_3$ and $\epsilon_4$, respectively. The exact values obtained were: final minimum growth rate 0.061416, $\epsilon_1=3.78794$, $\epsilon_2=3.37848$, $\epsilon_3=2.12919$, $\epsilon_4=0.704396$. This command can be modified to analyze situations with different budgets $B$ and stage-specific growth rates $r_i$ (or $k_i$) by substituting these into this command {\tt Minimize[\{5.47*(1 - r1 * (1 - Exp[-r1 x]))*(1 - r2 * (1 - Exp[-r2 y]))*(1 - r3*(1 - Exp[-r3 z]))*(1 - r4*(1 - Exp[-r4 w])), x + y + z + w == B \&\& x >= 0 \&\& y >= 0 \&\& z >= 0 \&\& w >= 0\},  \{x, y, z, w\}]}. It can also be modified to work with entirely different $p_i=p_i(\epsilon_i)$ functions by replacing the current ones, i.e. {\tt Minimize[\{5.47*(1 - p1(x))*(1 - p2(y))*(1 - p3(z))*(1 - p4(w)), x + y + z + w == B \&\& x >= 0 \&\& y >= 0 \&\& z >= 0 \&\& w >= 0\},  \{x, y, z, w\}]}. More stages can also be included by adding variables and inserting additional factors into the product that correspond to those stages.

\section*{Code Availability Statement}{All MATLAB code required to replicate the numerical results in this paper can be accessed here \url{https://github.com/danielstrombom/SLFOC}.


\bibliographystyle{unsrt}

\end{document}